\documentstyle[12pt]{article}
\newcommand{\sect}[1]{\section{#1}\setcounter{equation}{0}}

\def\Z{{\bf Z}}

\def\A{{\bf A}}
\def\D{{\bf D}}
\def\y{{\bf y}}
\begin{document}\bigskip
\hskip 3.7in\vbox{\baselineskip12pt
\hbox{NSF-ITP-96-54}\hbox{hep-th/9606165}}
\bigskip\bigskip\bigskip

\centerline{\large \bf Tensors from K3 Orientifolds}

\bigskip\bigskip

\centerline{\bf Joseph Polchinski} 
\medskip
\centerline{Institute for Theoretical Physics}
\centerline{University of California}
\centerline{Santa Barbara, CA\ \ 93106-4030}
\medskip
\centerline{e-mail: joep@itp.ucsb.edu}
\bigskip\bigskip

\begin{abstract}
\baselineskip=16pt
Recently Gimon and Johnson~\cite{GJ} and Dabholkar and Park~\cite{DP2}
have constructed Type I theories on K3 orbifolds.  The spectra
differ from that of Type I on a smooth K3, having extra tensors.  We
show that the orbifold theories cannot be blown up to smooth K3's, but
rather $\Z_2$ orbifold singularities always remain.  Douglas's recent
proposal to use D-branes as probes is useful in
understanding the geometry.  The
$\Z_2$ singularities are of a new type, with a different orientifold
projection  from those previously considered.  We also find a new
world-sheet consistency condition that must be satisfied by orientifold
models. 
\end{abstract}
\newpage
\baselineskip=18pt

\sect{A Puzzle}

Orientifolds are a generalization of orbifolds, allowing the construction
of interesting string theories based on free world-sheet fields. 
Although discovered some time ago~\cite{ori}, they have recently
attracted renewed interest because they are related by string dualities
to many other vacua.  In particular, there has been a series of papers
constructing models of this type in $d=6$ with $N=1$
supersymmetry~\cite{GP,DP,GJ,DP2}.\footnote{Orientifolds with $d=6$
$N=1$ supersymmetry were first found in refs.~\cite{sagten}.
It is likely
that some of the models found in these papers are identical to models in
refs.~\cite{GP,DP,GJ,DP2}.  However the puzzle we consider depends on the
spacetime picture, and so is not evident in these fermionic constructions.  For
further background on D-branes and orientifold see ref.~\cite{PCJ}.}  In
this note we would like to resolve a small puzzle arising from some of this
work.  This will lead us to a new orientifold construction and also to a
world-sheet consistency condition not previously noticed, though fortunately
satisfied by all the models of refs.~\cite{GP,DP,GJ,DP2}.  It also provides a
nice illustration of the recent idea of using D-branes as probes of spacetime
geometry~\cite{douglas}.

The puzzle is this.  Refs.~\cite{GP,GJ,DP2} all construct what appears to
be the Type I string on a $K3$ orbifold, in that one twists the IIB
theory on
$T^4$ by world-sheet parity $\Omega$ (producing the Type I string) and
by a spacetime $\Z_N$ rotation (producing K3).  For the $\Z_2$
case~\cite{GP}, the spectrum agrees with that of the Type~I string on a
smooth $K3$.\footnote
{The blowing up of the orbifold singularities was discussed in detail
in ref.~\cite{berk}.}  However, for $N \geq 3$ the orientifold spectrum
does not agree with that on smooth $K3$.  The latter, like its heterotic
dual, always has a single antisymmetric tensor multiplet, while in the
orientifold an extra $m$ tensors live at each $\Z_{2m+1}$ or $\Z_{2m+2}$
fixed point~\cite{GJ,DP2}.

The cause of the discrepancy is that the parity operator $\Omega$ of the
orientifold theory is not given by the limit of the $\Omega$ of the smooth
theory.\footnote{This is also under study by E. Gimon and
C. Johnson~\cite{GJ2}, by A. Dabholkar and J. Park~\cite{DP3}, and by J. Blum.}
Let us consider first the Type IIB theory at a $\Z_N$ orbifold
point (also known as an $A_{N-1}$ singularity).  The closed string Hilbert
space has twisted sectors labeled by
$k = 1, \ldots N-1$,
\begin{equation}
Z^{3,4}(2\pi) = \alpha^k Z^{3,4}(0),\qquad \alpha = e^{2\pi i/N},
\label{twist}
\end{equation}
where $Z^l = X^{2l} + i X^{2l+1}$.
The right-moving NS sector contains two states which are singlets under
the massless $SO(4)$ little group, while the right-moving R sector
contains a doublet.  The left-moving sectors are the same, so the full
spectrum is
\begin{equation}
(2\cdot{\bf 1}+{\bf 2}) \times (2\cdot{\bf 1}+{\bf 2})
= 5 \cdot {\bf 1} + {\bf 3}.
\end{equation}
These five singlets plus anti-self-dual tensor form a tensor multiplet of
Type IIB supergravity, call it ${\cal T}_k$, which decomposes to a tensor
multiplet and hypermultiplet of Type I\@.  Thus there are $N-1$ tensor
multiplets associated with the fixed point.  This is in agreement with the
limit of the smooth $K3$ spectrum.  At the fixed point there are $N-1$
collapsed two-spheres.  Each gives rise to a an anti-self-dual tensor
(from the self-dual four-form of the IIB theory), plus three moduli
from the metric and two theta parameters, from the R and NS two-forms.

Under the IIB parity operator $\Omega$, the metric and R two-form are
even and the four-form and the NS two-form are odd.  Projection onto
$\Omega = +1$ thus leaves 
$N-1$ hypermultiplets at the fixed point.  The parity operator
of the orientifold theory, call it
$\Omega'$, acts differently~\cite{GJ,DP2}.  Reversing the orientation of
the string changes the $\alpha^k$ twist~(\ref{twist}) to $\alpha^{-k} =
\alpha^{N-k}$. Except when $k = N-k$, this is off-diagonal and so one of
the two linear combinations ${\cal T}_k \pm {\cal T}_{N-k}$ is even and the
other odd.  For $k = N-k$ (so that $N$ is even), $\Omega'$ takes ${\cal
T}_{N/2}$ to itself, and the tensor and one NS scalar are odd and the
other four scalars even.  Forming the orientifold by projecting onto
$\Omega' = +1$ then leaves $m-1$ tensor multiplets and $m$ hypermultiplets
for
$\Z_{2m}$ and $m$ tensor multiplets and $m$ hypermultiplets for
$\Z_{2m + 1}$.  Only for $\Z_2$ does $\Omega = \Omega'$.

Evidently $\Omega' = \Omega J$ where $J$ is some
symmetry of the orbifold CFT which only acts on the fields near the
fixed point.  In the rest of the paper we will study this $J$
in more detail, not just in the $\Z_N$ orbifold limit but for the full
moduli space of $\Omega'$-invariant $K3$'s.  To make the discussion simple
we focus on a $\Z_3$ fixed point.  From the above discussion, this
has two $\Omega = +1$ hypermultiplets, only one of which has $\Omega' =
+1$.  Consider the ALE space, the local region of $K3$, obtained by
turning on the $\Omega' = +1$ moduli.  This cannot be smooth as
there would be no candidate for $J$, a symmetry which must act trivially
at long distance from the blown up fixed point.  But there is indeed a
family of singular ALE spaces which includes the $\Z_3$ orbifold and
which is parameterized by one hypermultiplet.  These spaces have $A_1$
singularities, $\Z_2$ orbifold points.

From the orbifold point of view it is hard to see a $\Z_3$ fixed point
deform into a $\Z_2$ fixed point, but from other points of view it is
simple.  The metric for a general $\Z_N$ ALE space is of the
form~\cite{ale}
\begin{eqnarray}
&&ds^2 = V^{-1}(dt - \A 
\cdot d\y)^2 + V
d\y \cdot d\y \nonumber\\
&&V = \sum_{i=0}^{N-1} \frac{1}{|\y - \y_i|},
\qquad \mbox{\boldmath$\nabla$}V = \mbox{\boldmath$\nabla$}
\times\A \label{alemet}
\end{eqnarray}
where $\y$ is a three-vector and $t$ has period
$4\pi$.  The $3(N-1)$ moduli mentioned earlier are just the $N-1$
differences $\y_i - \y_0$;
by a translation one can set $\y_0= 1$.  For $N=3$, the
$\Z_3$ orbifold singularity occurs when the three `charges' are
coincident,
$\y_0= \y_1=\y_2$.
Pulling one charge away and leaving two coincident
leaves a $\Z_2$ fixed point, and these singular ALE spaces are indeed
parameterized by one hypermultiplet.

There are no further hypermultiplets to associate with blowing up the
$\Z_2$ fixed point but there is a tensor multiplet, so this is
different from the $\Z_2$ fixed points of ref.~\cite{GP}.  The difference
is now clear: the $\Omega'$ projection is just keeping the opposite
states from $\Omega$ at the fixed point, so that $J$ is the $\Z_2$
symmetry which is $-1$ in the $\Z_2$-twisted sector and $+1$ in the
untwisted sector.\footnote
{This is after the $\Z_3$ singularity is partly resolved into a $\Z_2$
singularity.  On the original orbifold with $\Z_3$ singularity, $J$ must
interchange the sectors twisted by $\alpha$ and $\alpha^2$ while
leaving the untwisted sectors invariant.  This symmetry is not manifest,
for example it is not a symmetry of operator products involving $\Z_3$
non-invariant operators, but must be present in the orbifold CFT.}

In the next section we study $\Omega'$ orientifolds of free field
theory.  In the final section we verify the above argument about the ALE
geometry by using D-branes to measure the blown-up metric directly.  By
the same method we find that the generic blowup of the $\Z_{2m+1}$ and
$\Z_{2m+2}$ singularities of refs.~\cite{GJ,DP2} leaves $m$ separate
$A_1$ singularities, each with an associated tensor multiplet.

\sect{A New $\Z_2$ Orientifold}

Now let us study the new $\Z_2$ fixed point in isolation.  Start with the
IIB string in 10 dimensions.  Twist by a reflection $R$ of $X^{6,7,8,9}$
to produce an orbifold point at the origin, and then twist by $\Omega J$.
Note that this is akin to an asymmetric orbifold, in that the symmetry
$J$ does not exist until after the first twist.  Note also that if the
second twist is by $J$ rather than $\Omega J$ it simply undoes the
$R$-twist.  Projecting onto $J=+1$ removes the $R$-twisted states from
the spectrum, while the $J$-twisted states are by definition of $J$
those which have a branch cut relative to the $R$-twisted states---that
is, they are the $R$-odd states that were removed by the first
projection.

In order for the $\Omega J$ twist to be consistent, $\Omega J$ must be
conserved.  For purely closed string processes this follows because
$\Omega$ and $J$ are both conserved.  It is also necessary to
check the closed-to-open transition: $\Omega J$ as defined in the
closed string and open string sectors must be conserved by this
transition.  This must be true for all orbifold and orientifold twists in
open string theory, and is the missing consistency condition mentioned
earlier.  We will therefore also reexamine the earlier models.

To be specific we consider the transition between the closed string RR
ground state and the open string vector.  The analytic parts of the
relevant vertex operators are
\begin{eqnarray}
V_\alpha &=& e^{-\phi/2} S_\alpha \nonumber\\
T'_\alpha &=& e^{-\phi/2} S'_\alpha T \nonumber\\
V^\mu &=& e^{-\phi} \psi^\mu \label{verts}
\end{eqnarray}
for the untwisted R ground state, twisted R ground state, and untwisted
NS vector respectively.  Here $\phi$ is the bosonized ghost~\cite{fms},
$S_\alpha$ the spin field (spinor indices are ten-dimensional on
unprimed objects and six-dimensional on primed), and $T$ the
internal part of the twisted $R$ ground state.  All of these operators
are weight $(1,0)$, with the corresponding $(0,1)$ operators denoted by a
tilde. The relevant closed-to-open amplitudes on the unit disk are
determined purely by Lorentz invariance,
\begin{eqnarray}
\langle V_\alpha(0) \tilde V_\beta(0) V^\mu(1) \rangle &=&
\Gamma^\mu_{\alpha\beta} \nonumber\\
\langle T'_\alpha(0) \tilde T'_\beta(0) V^\mu(1) \rangle &=&
\Gamma'^\mu_{\alpha\beta} .  \label{amps}
\end{eqnarray}

Now consider the effect of orientation reversal $\Omega$.  This takes
\begin{eqnarray}
V_\alpha \tilde V_\beta &\to& \tilde V_\alpha V_\beta\ =
\ -V_\beta \tilde V_\alpha \nonumber\\
T'_\alpha \tilde T'_\beta &\to& \tilde T'_\alpha T'_\beta\ =
\ -T'_\beta \tilde T'_\alpha \nonumber\\
V^\mu &\to& -V^\mu.
\end{eqnarray}
Note that the R vertex operators anticommute, being spacetime spinors.
That the photon $V^\mu$ is $\Omega$-odd is familiar, though less obvious
in the $-1$ picture~(\ref{verts}) than in the 0 picture
where it is a tangent derivative.  The ten-dimensional $\Gamma$ matrices
are symmetric so the untwisted amplitude~(\ref{amps}) is even under
$\Omega$, but the six-dimensional $\Gamma$ matrices are antisymmetric and
the twisted amplitude is odd.  The full amplitude also contains a
Chan-Paton factor.  The twist $R$ acts on the Chan-Paton factors as a
matrix $\gamma_R$, so the Chan-Paton trace for the twisted amplitude
contains a factor $\gamma_R$: it is of the form
\begin{equation}
{\rm Tr}(\gamma_R \lambda_1 \ldots \lambda_n) .
\end{equation}
Parity $\Omega$ takes $\lambda_i \to (\gamma_\Omega^{-1} \lambda_i
\gamma_\Omega)^T$ and reverses the order $1, \ldots, n$.  This is
equivalent to $\gamma_R \to \gamma_\Omega \gamma_R^T \gamma_\Omega^{-1}$.
Conservation of $\Omega$ in the full amplitude then requires
\begin{equation}
\gamma_R = - \gamma_\Omega \gamma_R^T \gamma_\Omega^{-1}. \label{new}
\end{equation}
This was not imposed explicitly in ref.~\cite{GP}, though the condition
that $[\Omega, R] = 0$ hold on the Chan-Paton factors does imply
eq.~(\ref{new}) up to a sign.  In fact, the sign is correct for the model
of ref.~\cite{GP}.  It is also correct for the $\Z_2$ twist fields of
refs.~\cite{GJ,DP2}, because this sector is the same as in ref.~\cite{GP}
for all models.  For twists other than $\Z_2$ the new condition
is not as interesting.  For these, $\Omega$ takes the twist $g$ into a
different twist $g^{-1}$ so it governs which linear combination of the
two sectors appears.  This is necessary to get the correct vertex
operators, but does not affect the spectrum.  Similarly in ref.~\cite{DP},
$\Omega$ is replaced by an operator $\Omega S$ which acts off-diagonally
on the $R$-fixed points. 

Now consider the operator $\Omega J$.  By definition, this has an extra
minus sign in its action on $T'_\alpha \tilde T'_\beta$, so the above
argument leads to
\begin{equation}
\gamma_R = + \gamma_{\Omega J} \gamma_R^T \gamma_{\Omega J}^{-1}.
\label{new2} 
\end{equation}
In the nine-brane sector, cancellation of the ten-form tadpole
requires as always 32 nine-brane indices with $\gamma_{\Omega J,9} = 1$ in
an appropriate basis.  Then $\gamma_{R',9}$ is now symmetric, and so
is $\gamma_{\Omega J R,9} \ \propto\ \gamma_{R,9} \gamma_{\Omega J,9}$. 
The Chan-Paton algebra then implies that $\gamma_{\Omega J R,5}$ is
antisymmetric, the opposite of $\gamma_{\Omega R,5}$ in ref.~\cite{GP}. 
Examining the tadpoles of ref.~\cite{GP}, this changes the sign of the
cross-term in the untwisted six-form tadpole so that the fixed point has
the opposite charge from the usual $\Z_2$ fixed point: $+\frac{1}{2}$
times the five-brane charge for the new fixed point (call it type B)
versus $-\frac{1}{2}$ for the old (call it type A).

The vanishing of the twisted tadpole is not
automatic as it is at the usual fixed point.  At the latter the $\Omega$
projection removes the dangerous twisted sector six-form along with the
two-form, but the
$\Omega J$ projection leaves both.  So if for example there are no
five-branes at some fixed point, the tadpole condition of ref.~\cite{GP}
implies that Tr$(\gamma_{R,9}) = 0$ for the $\Omega J$ projection.  The
symmetry of
$\gamma_{R,9}$ and the surviving orthogonal change of basis after setting
$\gamma_{\Omega J,9} = 1$ then allow to take
\begin{equation}
\gamma_{R,9} = \left[ \begin{array}{cc} I_{16} & 0 \\ 0 & -I_{16}
\end{array} \right].  \label{gr9}
\end{equation}

We have studied the fixed point in a non-compact space but now let us
build a compact model.  The six-form charges found above do not allow
all sixteen fixed points to be of the new type as all six-form sources
would have the same sign, but they suggest a model with eight fixed
points of type A, eight of type B, and no five-branes.  
Indeed this
is possible.  Take
$K3$ to be a hypercube of side 2, so the coordinates the fixed point are
all 0 or 1.  Consider the eight fixed points with $X^6 = 0$.  The product
$J_0$ of the eight separate
$J$'s is not conserved.  However, the transition of a string from a fixed
point with $X^6 = 0$ to one with $X^6 = 1$ produces also a string with
winding number $w_6$ odd (in the orbifold $w_6$ is only defined mod~2), so
that $J_0 (-1)^{w_6}$ is conserved.  To see this, define $R$ as the
reflection which leaves $(0,0,0,0)$ invariant.  Then $(1,0,0,0)$ is left
invariant by $T_6 R$ where $T_6$ is a translation by 2 units in the $X^6$
direction.  A transition from a state of monodromy $R$ to one of
monodromy $T_6 R$ produces also a string of monodromy $T_6^{-1} \cong
T_6$, i. e., odd winding number.

Note that it is not possible in general to mix fixed points of
different types in an arbitrary way.  It is only consistent to project
(gauge) on an operation which is a symmetry of string theory.  One sees
from the above that fixed points of types A nd B can only be combined in
groups of eight.

The prescription then is to project the $K3$ orbifold of the
IIB theory by $\Omega J = \Omega J_0 (-1)^{w_6}$ and add in the usual 32
nine-brane indices.
The symmetries act on the Chan-Paton
matrices as
$\gamma_{\Omega J} = 1$ and $\gamma_R$ given by eq.~(\ref{gr9}),
while the condition that the eight fixed points with $X^6 = 1$ be of
type A require that $\gamma_{T_6 R}$ be as in
ref.~\cite{GP},
\begin{equation}
\gamma_{T_6 R} = \left[ \begin{array}{cc} 0 & I_{16} \\ - I_{16} & 0
\end{array} \right].  \label{gdr9}
\end{equation}
This implies a nontrivial Wilson line $\gamma_{T_6} = \gamma_{T_6 R}
\gamma_{R}^{-1}$.  The resulting model satisfies all algebraic and
tadpole conditions. 

The untwisted closed string sector contributes the usual gravitational
multiplet, tensor multiplet, and four hypermultiplets.  The twisted
sectors obviously contribute eight tensor multiplets and eight
hypermultiplets.  The open string gauge group is broken from $SO(32)$
to $SO(16) \times SO(16)$ by $\gamma_R$ and then to $SO(16)$ by
$\gamma_{T_6}$.  The open string spectrum also includes a hypermultiplet
in the adjoint of $SO(16)$.

This is the same spectrum found in the models of ref.~\cite{DP}.  Those
models had five-branes but no
nine-branes.  Not surprisingly then, the model found here is equivalent
to those under $T$-duality on the $X^{6,7,8,9}$ axes. The
$T$-duality acts on the fixed point states as a discrete Fourier
transform, taking a fixed point with coordinates $X^m$ into a linear
combination of all fixed points, the fixed point with coordinates $Y^m$
being weighted by $2^{-4/2} (-1)^{X \cdot Y}$.  It follows that $J_0 =
-(-1)^{X^6}$ maps to the translation $Y^6 \to Y^6 + 1$ of the fixed
point states, up to a sign that can be absorbed in the definition of the
states.  The factor $(-1)^{w_6}$ maps to $(-1)^{n_6}$, which translates
untwisted states by half the lattice spacing.  Thus $J_0 (-1)^{w_6}$ maps
to the full operator $T_6^{1/2}$ for translation by half the lattice
spacing. In the notation of ref.~\cite{DP}, $T_6^{1/2} = RS$ and the
orientifold groups
$\{ 1, R, \Omega S, \Omega RS\}$ of the two models are the same.  Since
$R$ here maps to $R$ of ref.~\cite{DP}, our model is dual
to symmetric solution of that paper.  But if we translate our model 
$X^6 \to X^6 + 1$ and then take the $T$-dual, $T_6 R$ of our model would
map to $R$ of that paper, giving the antisymmetric solution.  
It must be that the two solutions of ref.~\cite{DP} are equivalent under
a redefinition by the image of $T_6^{1/2}$, namely $J_0 (-1)^{w_6}$. 
One finds that this in indeed the case.\footnote{In verifying this,
note the in models with five-branes only, open strings should be regarded
as having winding number $w_6 = (X^6(\pi) - X^6(0))/2\pi R_6$.}

The models of ref.~\cite{DP} also have two kinds of fixed point, neither
of which is A or B.  Half the fixed points are ordinary orbifold points,
fixed by $R$ but no operation involving $\Omega$.  These fixed points
have no six-form charge, there being no associated crosscap, and as
discussed in the introduction have a tensor and a hypermultiplet.  The
other type are fixed only by an operation $\Omega S$, equivalent to
$\Omega R$.  These must have charge $-1$, by overall neutrality of the
model, and have no associated twisted states. 
Finally, Dabholkar and
Park have recently found yet another kind of $\Z_2$ singularity in
orientifold models, having a tensor multiplet but six-form
charge $-\frac{1}{2}$.  This is based on the projection
$\Omega J$ but with a different action on the Chan-Paton factors.  In
particular, the operator $\Omega^2$ is $+1$ in the 59 open string
sector, having the minus sign noted in ref.~\cite{GP} plus an additional
minus sign because the 59 sector has half-integer rather than integer
masses.

\sect{ALE Geometry}

We can directly test our argument about the geometry of the
$\Omega'$-invariant $K3$'s by using the D-string as a probe, as recently
proposed by Douglas~\cite{douglas}.  All results in the present section
are already implicit in refs.~\cite{kron,comm,quiv}, but the D-probe idea
seems very promising and so it is worthwhile to work out this simple
example explicitly.

We start with the $\Z_2$ ALE space.  Consider the IIB theory with a
$\Z_2$ orbifold point (or more precisely six-plane) at $X^{6,7,8,9} = 0$,
and add a D-string in this plane at $X^{2,\ldots,9} = 0$.  In order for
the string to be able to move off the fixed plane it needs two
Chan-Paton indices, for the string and its $\Z_2$ image.  Since $R$
takes the D-string into its image, $\gamma_R$ is the Pauli matrix
$\sigma^1$.  The massless NS spectrum of the string is then
(in terms of the 0 picture vertex operators)
\begin{eqnarray}
&&\partial_t X^\mu \sigma^{0,1}, \qquad \mu = 0,1 \nonumber\\
&&\partial_n X^i \sigma^{0,1}, \qquad i = 2,3,4,5 \nonumber\\ 
&&\partial_n X^m \sigma^{2,3}, \qquad m = 6,7,8,9.
\end{eqnarray}
These are respectively a gauge field, the position of the string within
the six-plane, and the transverse position.  Call the corresponding
D-string fields $A^\mu, x^i, x^m$, all $2\times2$ matrices.  The
bosonic action is the $d = 10$ $U(2)$ Yang-Mills action, dimensionally
reduced and $R$-projected (which breaks the gauge symmetry to $U(1)
\times U(1))$.  In particular, the potential is
\begin{equation}
U = 2\sum_{i,m} {\rm Tr}([x^i,x^m]^2) + \sum_{m,n} {\rm
Tr}([x^m,x^n]^2). \label{pot}
\end{equation}
The moduli space thus has two branches.  On one, $x^m = 0$ and $x^i =
u^i \sigma^0 + v^i \sigma^1$.  This corresponds to two D-strings moving
independently in the plane, with positions $u^i \pm v^i$.  The gauge
symmetry is unbroken, giving independent $U(1)$'s on each
D-string.\footnote
{These D-strings can actually be regarded as collapsed three-branes
wrapped on the orbifold point.  They couple to the corresponding RR
field, the twisted-sector tensor.  Because the $\theta$-parameter from
the NS sector is nonzero~\cite{asp}, they also carry the untwisted
six-form charge.  When the theta parameter is tuned to zero these
strings become tensionless~\cite{comm}.} 
On the other branch, $x^m$ is nonzero and $x^i = u^i
\sigma^0$.  The $\sigma^1$ gauge invariance is broken and so by gauge
choice $x^m = w^m \sigma^3$.  This corresponds to the D-string moving
off the fixed plane, the string and its image being at
$(u^i, \pm w^m)$.

Now let us turn on twisted-sector moduli.  Define complex $q^m$ 
by $x^m = \sigma^3 {\rm Re}(q^m) + \sigma^2 {\rm Im}(q^m)$, and define two
doublets,
\begin{equation}
\Phi_0 = \left( \begin{array}{c} q^6 + i q^7 \\ q^8 + i q^9
\end{array} \right), \qquad
\Phi_1 = \left( \begin{array}{c} \bar q^6 + i \bar q^7 \\ \bar q^8 + i
\bar q^9
\end{array} \right).
\end{equation}
These have charges $\pm 1$ respectively under the $\sigma^1$ $U(1)$.
The three NSNS moduli can be written as a vector $\D$,
and the potential is proportional to
\begin{equation}
(\Phi_0^\dagger \mbox{\boldmath$\tau$} \Phi_0 - \Phi_1^\dagger
\mbox{\boldmath$\tau$} \Phi_1 + \D)^2,
\end{equation}
where the Pauli matrices are now denoted $\tau^a$ to emphasize that they
act in a different space.  This reduces to the second term of the
earlier potential~(\ref{pot}) when $\D = 0$.  Its form
is determined by supersymmetry, and the trilinear coupling between the
twisted sector field and two open string fields was demonstrated in the
appendix to ref.~\cite{quiv}.

For $\D \neq 0$ the
orbifold point is blown up.  The moduli space
of the D-string is simply the set of possible locations, that is, the
blown up ALE space.\footnote
{Note that the branch of the
moduli space with $v^i \neq 0$ is no longer present.}  The $z^m$ contain
eight scalar fields.  Three are removed by the $\D$-flatness condition,
that the potential vanish, and a fourth is a gauge degree of freedom,
leaving the expected four moduli.  In terms of supermultiplets, the
system has the equivalent of $d=6$ $N=1$ supersymmetry.  The D-string
has two hypermultiplets and two vector multiplets, which
are Higgsed down to one
hypermultiplet and one vector multiplet. 

The idea of
ref.~\cite{douglas} is that the metric on this moduli space, as seen in
the kinetic term for the D-string fields, should be the smoothed ALE
metric.  It is straightforward to verify this.  Define
\begin{equation}
\y = \Phi_0^\dagger \mbox{\boldmath$\tau$} \Phi_0.
\label{hopf}
\end{equation}
This gives three
coordinates on moduli space.  The fourth coordinate $t$ can be defined
\begin{equation}
t = 2\arg(\Phi_{0,1} \Phi_{1,1}).
\end{equation}
The period of $t$ is $4\pi$ because of the orbifold projection.
The $\D$-flatness condition
implies that
\begin{equation}
\Phi_1^\dagger \mbox{\boldmath$\tau$} \Phi_1 = \y
+ \D,
\end{equation}
and $\Phi_0$ and $\Phi_1$ are determined in terms of $\y$ and $t$, up to
gauge choice.

The original metric is $d\Phi_0^\dagger d\Phi_0 + d\Phi_1^\dagger
d\Phi_1$, but we need to
project this into the space orthogonal to the
$U(1)$ gauge transformation.\footnote{This whole construction,
imposing the \D-flatness conditions and making the gauge
identification, is known as the hyper-K\"ahler quotient~\cite{kron}.}  The
result is 
\begin{equation}
ds^2 = d\Phi_0^\dagger d\Phi_0 + d\Phi_1^\dagger
d\Phi_1 - \frac{(\omega_0 + \omega_1)^2}{4 (\Phi_0^\dagger \Phi_0 + 
\Phi_1^\dagger \Phi_1)}
\end{equation}
where
\begin{equation}
\omega_i = i(\Phi_i^\dagger d\Phi_i - d\Phi_i^\dagger \Phi_i)
\end{equation}.

It is now straightforward to express the metric in terms of $\y$ and $t$
using the identity $(\alpha^\dagger \tau^a \beta)(\gamma^\dagger \tau^a
\delta) = 2(\alpha^\dagger \delta)(\gamma^\dagger \beta) -
 (\alpha^\dagger \beta)(\gamma^\dagger \delta)$ for arbitrary doublets
$\alpha, \beta, \gamma, \delta$.
This implies, for example,
\begin{eqnarray}
&&\Phi_0^\dagger \Phi_0 = |\y|,\qquad \Phi_1^\dagger \Phi_1=|\y + \D|,
\nonumber\\ &&d\y \cdot d\y\ =\ |\y| d\Phi_0^\dagger d\Phi_0 - \omega_0^2
\ =\ |\y + \D| d\Phi_1^\dagger d\Phi_1 - \omega_1^2,
\end{eqnarray}
and the metric is readily found to be of the ALE form~(\ref{alemet})
with $\y_0 = 0$, $\y_1 = \D$, up to a normalization that can be absorbed
in a coordinate transformation.  In particular, the vector potential is
\begin{equation}
\A(\y) \cdot d\y = |\y|^{-1} \omega_0 + |\y + \D|^{-1} \omega_1 + dt,
\end{equation}
and the field strength is readily obtained by taking the exterior
derivative and using the identity $\epsilon^{abc} (\alpha^\dagger \tau^b
\beta)(\gamma^\dagger \tau^c\delta) = i (\alpha^\dagger \tau^a
\delta)(\gamma^\dagger \beta) - i(\alpha^\dagger \delta)(\gamma^\dagger
\tau^a \beta)$.

This all extends to the $\Z_N$ case.  In order to move away from
the fixed point the D-string needs $N$ Chan-Paton indices, with the $Z_N$
matrix $\gamma_\alpha = a$ taking each index into the next.  Define
another $N\times N$ matrix $b$ with the properties $ab = \alpha ba$
($\alpha = e^{2\pi i/N}$), $b^N = 1$, which together with $a^N$ define
$a$ and $b$ up to change of basis.  The open string states in a
convenient basis are
\begin{eqnarray}
&&\partial_t X^\mu \lambda,\quad a^{-1} \lambda a = \lambda
\ \ \Rightarrow\ \ \lambda = P_r \nonumber\\
&&\partial_n X^i \lambda,\quad a^{-1} \lambda a = \lambda 
\ \ \Rightarrow\ \ \lambda = P_r \nonumber\\ 
&&\partial_n Z^l \lambda,\quad a^{-1} \lambda a = \alpha^{-1} \lambda
\ \ \Rightarrow\ \ \lambda = b P_r \nonumber\\
&&\partial_n \bar Z^l \lambda,\quad a^{-1} \lambda a = \alpha
\lambda\ \ \Rightarrow\ \ \lambda = b^{-1} P_r 
\end{eqnarray}
where
\begin{equation}
P_r = N^{-1/2} \sum_{k=0}^{N-1} \alpha^{rk} a^k
\end{equation}
are projection operators.
Call the corresponding fields $A^\mu_r$, $x^\mu_r$, $\Phi^l_r$,
$\bar \Phi^l_r$ with $r = 0, \ldots, N-1$.  With the lower index
suppressed these are $N\times N$ matrices.  The covariant derivative is
\begin{equation}
D_\mu \Phi = \partial_\mu \Phi + i [ A_\mu, \Phi ] .
\end{equation}
Noting that $P_r b = b P_{r+1}$, this implies
\begin{equation}
D_\mu \Phi_r = \partial_\mu \Phi_r + i (A_{r-1,\mu} - A_{r,\mu}) \Phi_r,
\end{equation}
with $r = N \equiv r = 0$.
The
$N$ $U(1)$ \D-flatness conditions require 
\begin{equation}
\Phi_{r+1}^\dagger \mbox{\boldmath$\tau$} \Phi_{r+1} -\Phi_r^\dagger
\mbox{\boldmath$\tau$} \Phi_r = \D_r.
\end{equation}
Note that this implies that there are $N-1$ \D-terms, as the sum must
vanish.  The metric is then readily written in ALE form, with  
\begin{equation}
\y_i = \sum_{r = 0}^{i-1} \D_r.
\end{equation}

Now we can return to our original purpose, which was to identify the 
$\Omega'$-invariant ALE spaces.  Recall above that the matrix $a$ is
interpreted as connecting each D-string with its image rotated by
$\alpha$.  A Chan-Paton factor proportional to $a^k$ is then an open
string with one end at one image and the other rotated by $\alpha^k$. 
Orientation reversal, whether $\Omega$ or $\Omega'$, switches the
endpoints and so takes this into a string with Chan-Paton factors
$a^{-k}$.  In terms of the projection operators this is $P_r \to
P_{N-r}$, and so
\begin{equation}
\Omega':\quad \Phi_r \to \Phi_{N-r}, \quad \D_r \to -\D_{N-r-1}.
\end{equation}
It follows that for $N=3$, requiring the twisted background to be
$\Omega'$-even implies that $\D_1 = 0$, $\D_2 = - \D_0$, and so $\y_1 =
\y_2 \neq \y_0$ as conjectured.  Similarly for the general $Z_N$ fixed
point, one finds that $\y_i = \y_{N-i}$, leaving $m$ collapsed two-spheres
for $N = 2m+1$ or $2m+2$.

It would be interesting to understand the strong-coupling behavior of
the theory near the ALE singularity.  Away from the singularity it is strongly
coupled Type I and so weakly coupled heterotic $SO(32)$, but there is no
perturbative background of the heterotic string with extra tensors.  In this
connection we should note that there have been many recent discussions of extra
tensors in the contexts of M-theory and F-theory; it is not clear whether these
are directly relevant, since the {\it local physics} near but not at the
singularity is just the heterotic string.

\subsection*{Acknowledgments} 

I would like to thank Atish Dabholkar, Eric Gimon, Clifford Johnson, and
Jaemo Park for extensive discussions of their models, and
Hirosu Ooguri for a discussion.  This work is supported by NSF
grants PHY91-16964 and PHY94-07194.


\begin{thebibliography}{99}
\baselineskip=17pt

\bibitem{GJ} E. G. Gimon and C. V. Johnson, {\it ``K3
Orientifolds,''} preprint NSF-ITP-96-16, hep-th/9604129.

\bibitem{DP2} A. Dabholkar and J. Park, {\it ``Strings on
Orientifolds,''} preprint CALT-68-2051, hep-th/9604178.

\bibitem{ori}
A. Sagnotti, in Cargese '87, ``Non-perturbative Quantum
Field Theory,'' ed. G. Mack et. al. (Pergamon Press, 1988) p. 521;\\
V. Periwal, unpublished;\\
G. Pradisi and A. Sagnotti, Phys. Lett. {\bf B216} (1989) 59;\\
P. Horava, Nucl. Phys. {\bf B327} (1989) 461;\\
J. Dai, R. G. Leigh, and J. Polchinski,
Mod. Phys. Lett. {\bf A4} (1989) 2073. 

\bibitem{GP} E. G. Gimon and J. Polchinski, {\it ``Consistency Conditions
for Orientifolds and D-Manifolds,''} preprint NSF-ITP-96-01,
hep-th/9601038.

\bibitem{DP} A. Dabholkar and J. Park, {\it ``An Orientifold of Type IIB
Theory on K3,''} preprint CALT-68-2038, hep-th/9602030.

\bibitem{sagten}
M. Bianchi and A. Sagnotti, Phys. Lett. {\bf B247} (1990) 517;
Nucl. Phys. {\bf B361} (1991) 519;\\
A. Sagnotti, Phys. Lett. {\bf
B294} (1992) 196.

\bibitem{PCJ} J. Polchinski, S. Chaudhuri, and C. V. Johnson, {\it ``Notes
on D-Branes,''} preprint NSF-ITP-96-003, hep-th/9602052.

\bibitem{douglas} M. Douglas, {\it ``Gauge Fields and D-Branes,''}
M. R. Douglas, preprint RU-96-41, hep-th/9604198.

\bibitem{berk} M. Berkooz, R. G. Leigh,
J. Polchinski, J. H. Schwarz, 
N. Seiberg, and E. Witten, {\it ``Anomalies, Dualities and Topology in
$D=6$
$N=1$ Superstring Vacua,''} preprint RU-96-16,
NSF-ITP-96-21, CALT-68-2057, IASSNS-HEP-96/53, hep-th/9605184. 

\bibitem{GJ2} E. G. Gimon and C. V. Johnson, {\it ``Multiple Realisations of
$N=1$ Vacua in Six Dimensions,''} in preparation.

\bibitem{DP3} A. Dabholkar and J. Park, in preparation.

\bibitem{ale} T. Eguchi and A. J. Hanson, Ann. Phys. {\bf 120} (1979)
82;\\
G. W. Gibbons and S. W. Hawking, Comm. Math. Phys. {\bf 66} (1979) 291.

\bibitem{fms} D. Friedan, E. Martinec, and S. Shenker,
Nucl. Phys. {\bf B271} (1986) 93.

\bibitem{kron} A. Hitchin, A. Karlhede, U. Lindstrom, and M. Ro\v cek,
Comm. Math. Phys. {\bf 108} (1987) 535;\\
P. Kronheimer, J. Diff. Geom. {\bf 28} (1989) 665; {\bf 29} (1989) 685.

\bibitem{comm} E. Witten, {\it ``Some Comments on String Dynamics,''}
preprint IASSNS-HEP-95-63, hep-th/9507121.

\bibitem{quiv} M. R. Douglas and G. Moore, {\it ``D-Branes, Quivers, and
ALE Instantons,''} preprint RU-96-15, hep-th/9603167.

\bibitem{asp} P. S. Aspinwall, Phys. Lett. {\bf B357} (1995) 329, 
hep-th/9507012.


\end{thebibliography}
\end{document}